\documentclass[pra,twocolumn,notitlepage,longbibliography]{revtex4-1}
\usepackage{amsmath}
\usepackage{amssymb}

\usepackage{hyperref}
\usepackage{bm}
\usepackage{epsfig}

\usepackage[normalem]{ulem}

\newcommand {\diag}{\mathop{\mathrm{diag}}\nolimits}

\renewcommand {\Im}{\mathop\mathrm{Im}\nolimits}

\renewcommand {\phi}{{\varphi}}
\newcommand {\rmi}{{\rm i}}
\newcommand {\rmd}{{\rm d}}

\newcommand {\e}{{\rm e}}
\newcommand {\eps}{\varepsilon}

\begin{document}
\title{Quasi-flat  band  enables subradiant two-photon bound states }

\author{Alexander N. Poddubny}
\email{poddubny@coherent.ioffe.ru}
\affiliation{Ioffe Institute, St. Petersburg 194021, Russia}

\begin{abstract}
We study theoretically  the radiative lifetime of bound two-particle excitations in a  waveguide with an   array of two-level atoms, realising a 1D Dicke-like model. Recently, Zhang  et al.
\cite{Zhang2019arXiv} have numerically found  an unexpected sharp maximum of the bound pair lifetime when the array period $d$ is equal to $1/12$th of the light wavelength $\lambda_0$ [\href{https://arxiv.org/abs/1908.01818}{arXiv:1908.01818}]. 
We uncover  a rigorous transformation from the non-Hermitian Hamiltonian with the long-ranged  radiative coupling  to the nearest-neigbor coupling model with the radiative losses only at the edges.  This naturally explains the puzzle of long lifetime: the  effective mass of the bound photon pair also diverges for $d=\lambda_0/12$, hampering an escape of photons through the edges.  
We also  link the  oscillations of the lifetime with the number of atoms  to the nonmonotous quasi-flat-band dispersion of the bound  pair.
\end{abstract}
\date{\today}

\maketitle

\section{Introduction}

The array of atoms coupled to freely propagating photons is a paradigmatic system for quantum optics, well known at least since the work of Dicke in 1954~\cite{Dicke1954,KimbleRMP2018,Roy2017,Yudson1984}. However, recent technological advances with the cold atom~\cite{Corzo2019}
and superconducting qubit~\cite{vanLoo2013,Mirhosseini2019,Wang2019} systems have revived and boosted interest to this problem
 underlining its importance for future quantum technologies~\cite{Rabl2019,Masson2019,Kornovan2019,yuan2019}. Specifically, it has been understood that the new collective many-body effects emerge   when the distance between the atoms is varied. The physics of single-excited states is relatively straightforward: in the subwavelength  case there exist multiple strongly subradiant  
modes with the radiative lifetime scaling  as $N^3/d^2$ with the number of atoms $N$ and the spacing $d$~\cite{Albrecht2019}.
However, the two-particle subradiant excitations appear to be significantly  more complex due to the photon blockade that forbids double excitation of a single atom.
In particular, for small array periods $d\ll \lambda_0 $ the most subradiant two-particle states are fermionized hard-core bosons~\cite{Molmer2019} (here $\lambda_0=2\pi c/\omega_0$ is the light wavelength at the atom frequency).
 Interaction between strongly subradiant and less subradiant single-particle states gives rise 
to unusual  states when one photon is a standing wave and a second one is localized in the nodes of this wave or vice versa~ \cite{Zhong2019}. Bound two-photon states have also been predicted \cite{Zhang2019arXiv}. Interestingly and very unexpectedly, the lifetime of bound states  depends nonmonotonously on the array period. While one could expect a monotonous  decrease as $(\lambda_0/d)^2$, 
similar to the single-photon case, the lifetime  shows instead a sharp maximum for a  ``magic'' period $d=\lambda_0/12\ll \lambda_0$. 
Moreover, in the vicinity of the magic period the lifetime demonstrates an unusual dependence on the number of atoms. Instead of a monotonous increase $\propto N^3$, as in the single-particle case, it oscillates with the number of atoms. Despite the detailed numerical analysis in
Ref.~\cite{Zhang2019arXiv} the  physical origin of the unusual lifetime behavior in the vicinity of the magic period $d=\lambda_0/12$ remains unclear to the best of our knowledge. Here we reveal the connection between the numerical results of Ref.~\cite{Zhang2019arXiv} for the finite arrays and the center-of-mass dispersion of the bound pairs in the infinite structure.  We demonstrate, that the effective mass of the bound photon pair diverges for $d=\lambda_0/12$ and analyze how this affects the radiative escape of the photons and the resulting lifetime of the bound state.
 These results provide  a simple heuristic  recipe to protect the quantum  many-body correlations against the radiative losses.

\section{Diverging effective mass}
\begin{figure}[!b]
\centering\includegraphics[width=0.48\textwidth]{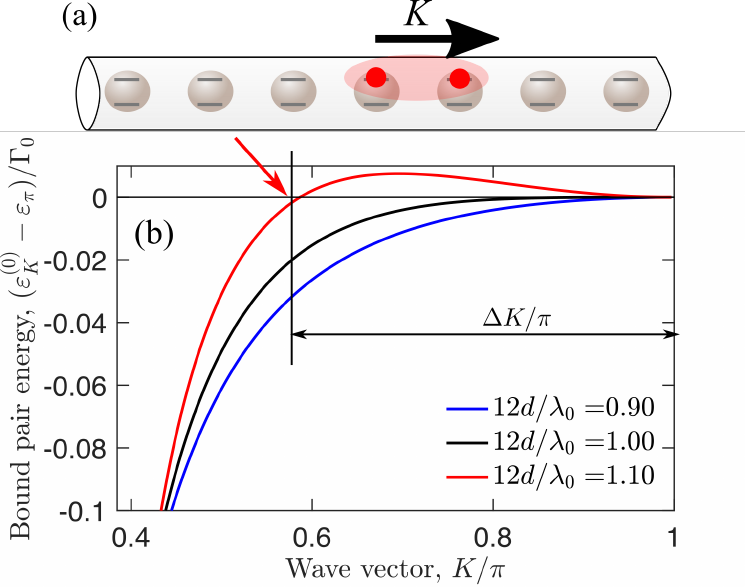}
\caption{(a) Schematic illustration of a bound two-particle state. The state propagates in an array of qubits in a waveguide with the center-of-mass wave vector $K$. (b) Energy dispersion of the bound state for three array periods, close to $\lambda/12$.
An arrow in (b) shows the point $K=\pi-\Delta K$, where $\eps_K=\eps_\pi$  for $d>\lambda_0/12$.
}
\label{fig:1}
\end{figure}
The system under consideration is shown in Fig.~\ref{fig:1}(a). It consists of $N$ periodically spaced identical two-level atoms. We are interested in the  two-particle excitations of this array. In the Markovian approximation they can be found from the Schr\"odinger equation 
~\cite{Ke2019arXiv}
\begin{equation}\label{eq:Sch}
\sum\limits_{r's'}H_{rs;r's'}\Psi_{r's'}=2\eps\Psi_{rs}\:.
\end{equation}
Here, the wave function $\Psi_{rs}$ describes the two-particle state $\sum_{r,s=1}^N\Psi_{rs} b^\dag_{r}b^\dag_{s}|0\rangle$. The indices $r$ and $s$ label the atoms and the creation operator  $b^\dag_{r}$  describes excitation of a given atom. 
The two-photon Hamiltonian reads
\begin{equation}
H_{rs;r's'}=H_{0,rr'}\delta_{ss'}+H_{0,ss'}\delta_{rr'}\:,\label{eq:H}
H_{0,rr'}=-\rmi \Gamma_{0}
\e^{\rmi \varphi |r-r'|}
\end{equation}
where $\Gamma_0$ is the radiative decay rate of an individual atom and $\phi=\omega_0d/c$ is the phase gained by light when passing between two neighbouring qubits. The energy $2\omega_0$ has been subtracted to shorten the notation, the total energy of two-photon excitation is $2\omega_0+2\eps$.
Equation Eq.~\eqref{eq:Sch} should be solved for symmetric bosonic excitations $\Psi_{rs}=\Psi_{sr}$ with the additional condition $\Psi_{rr}=0$, forbidding double excitation of a single two-level atom. 
Given the condition $\Psi_{rr}=0$, the equation can be also rewritten in a matrix form as \cite{Ke2019arXiv,Zhong2019}
\begin{equation}\label{eq:Schb}
H_0^{}\Psi+ \Psi H_0^{}-2\diag[ \diag ( H_0 \Psi)]=2\eps \Psi \:.
\end{equation}
The eigenstates of Eqs.~\eqref{eq:Sch},\eqref{eq:Schb} for finite $N$ have complex energies, with the imaginary part of energy determining  their radiative decay rate.

In addition to the solutions of Eq.~\eqref{eq:Sch} for a finite number of atoms we are also interested in the eigenstates of an infinite periodic array. In this case the pair of photons is characterized by the center-of-mass wave vector $K$ and the two-photon state can be written as~\cite{Zhang2019arXiv}
\begin{equation}
\Psi_{rs}=\e^{\rmi K(r+s)/2}\Phi_{r-s},\quad \Phi_{0}=0,\quad \Phi_{r}=\Phi_{-r}\:.\label{eq:com}
\end{equation}
Substituting Eq.~\eqref{eq:com} into Eq.~\eqref{eq:Sch}  we obtain the following system of equations describing the relative motion
of a pair of excitations
\begin{gather}\label{eq:relative}
\sum\limits_{s=-\infty}^{\infty}\mathcal H_{r,s}(K)\Phi_s=\eps_K \Phi_{r}\:,\Phi_0=0,\\
\mathcal H_{r,s}(K)=-\rmi\Gamma_{0}\cos \tfrac{K(r-s)}{2}\e^{\rmi\varphi|r-s|}\:.
\end{gather}
The system Eq.~\eqref{eq:relative} for a given center-of-mass wave vector has both continuous spectrum and  bound eigenstates with  discrete energies. The states of continuous spectrum are just scattering states of two quasi-independent polaritonic  excitations. Since there exist two single-particle polaritonic branches, the upper one and the lower one, there are three bands of two-particle scattering states in total arising from different combinations of two single-particle bands. In addition to the scattering state, there also exists a bound two-particle solution of Eq.~\eqref{eq:relative}, where the relative motion wave function $\Psi_m$ decays for $m\to\infty$.

\begin{figure}[!t]
\centering\includegraphics[width=0.48\textwidth]{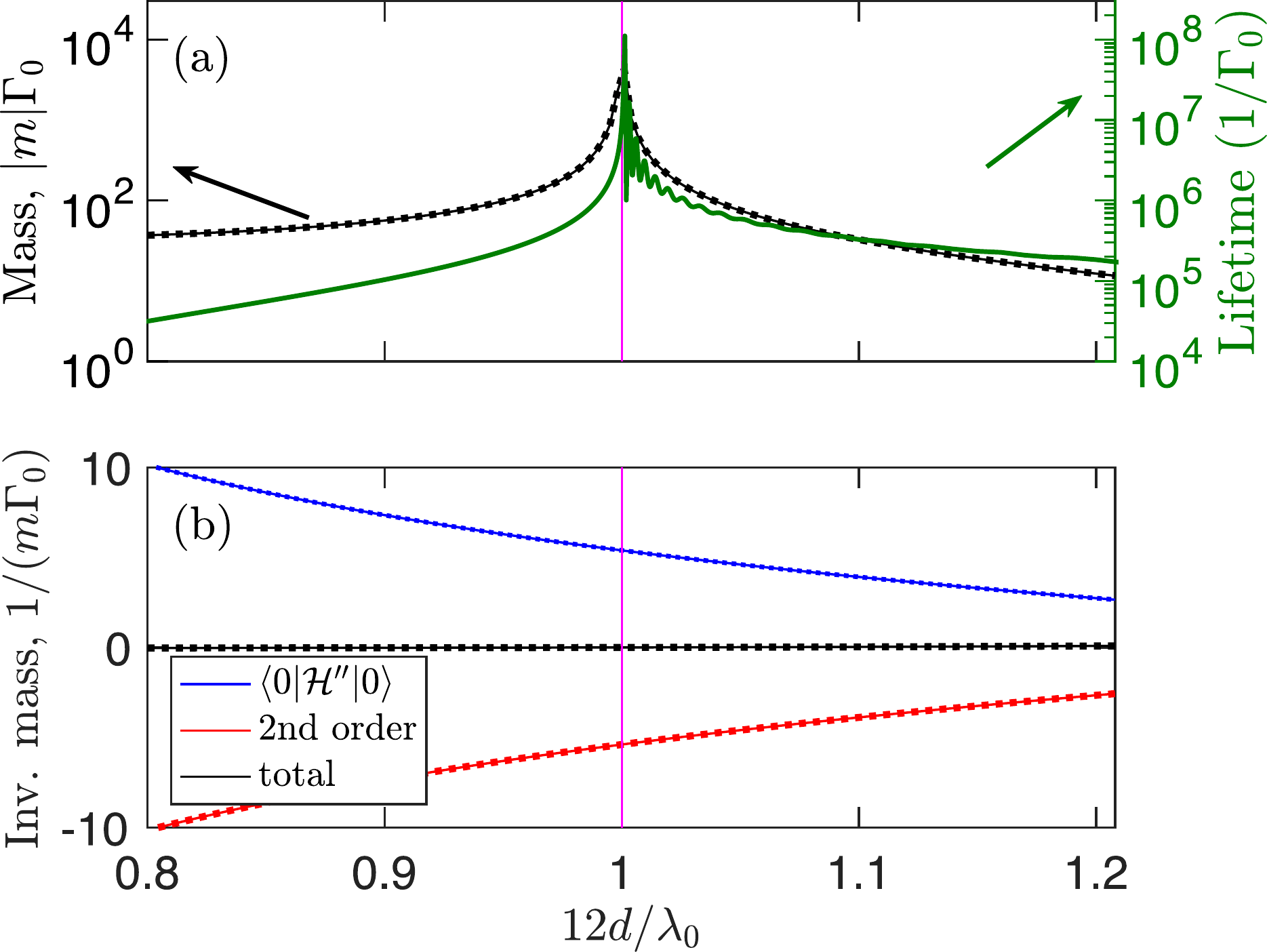}
\caption{(a) Left axis: Dependence of the effective mass of the photon pair at the point $K=\pi$ on the period of the array.
Lines are calculated numerically, dots present the analytical result Eq.~\eqref{eq:mass}.
Right axis: dependence of the longest lifetime of the  bound photon pair in the array with $N=80$ atoms on the array period.
(b) Contributions of the first and second terms from Eq.~\eqref{eq:kp} to the inverse effective mass.
}
\label{fig:2}
\end{figure}

The corresponding dispersion branch, calculated numerically from Eq.~\eqref{eq:relative}, is illustrated in Fig.~\ref{fig:2} for three different periods of the array $d$ close to $\lambda_0/12\equiv 6\pi c/\omega_0$. The calculation demonstrates that the curvature of the dispersion curve at the point $K=\pi$ changes its sign for $d=\lambda_0/12$. The numerical results indicate that the value $d=\lambda_0/12$ corresponds to the infinite mass at the Brillouin zone edge,  $\rmd^2\eps/\rmd K^2=0$ for $K=\pi$. As such, the dispersion of a bound pair becomes quasi-flat in a relatively wide range of the Brillouin zone.

Unfortunately, we were not able to obtain a close-formed analytical solution for the bound pair dispersion $\eps_K^{(0)}$ for arbitrary center-of-mass wave vectors. However, the behaviour near the extremum $K=\pi$  can be still analysed by means of usual $\bm k\cdot \bm p$ perturbation theory~\cite{Cardona}.
Namely, the bound pair is described by \cite{Zhang2019arXiv}
\begin{equation}\label{eq:bound}
 \Phi_{\pm 2r}^{(0)}=
(-1)^r\e^{-(r-1)\kappa}\sqrt{1-e^{-2\kappa}},r=1,2,\ldots
\end{equation}
where 
\begin{equation}\label{eq:kappa}
\kappa=-2\ln \cos 2\phi
\end{equation}
is the inverse effective  size of the bound pair. The bound pair has the energy
$\eps^{\rm (0)}_\pi=2\Gamma_0 \cot 2\phi$.
The scattering states for $K=\pi$ can be presented as
\begin{equation}\label{eq:scat}
 \Phi_{\pm (2r-1), q}=\sqrt{2}\cos q(r+\tfrac{1}{2})\:,
\end{equation}
where $r=1,2,\ldots$ and have the energy
\begin{equation}
\eps(q)=\frac{\Gamma_0\sin\phi\cos\phi}{\sin^2\phi-\cos^2\frac{q}{2}}\:.
\end{equation}
Here, $-\pi<q\le \pi$ is the wave vector of relative motion of the two particles. For small absolute values of $q$ we find $\eps(q)<0$, corresponding to the two excitations of the lower polaritonic branch. Large values of $q$, close to the Brillouin zone edge, have $\eps(q)>0$ and correspond to a double excitation of the upper polaritonic branch.

The effective mass of the bound pairs at $K=\pi$ reads 
\begin{equation}
\frac1{m}=
 \langle 0|\mathcal H''|0\rangle+
 {\rm v.p.}\int\limits_{-\pi}^{\pi} \frac{\rmd q}{2\pi} \frac{2\bigl| \langle 0|\mathcal H'|q\rangle\bigl|^2 }{\eps^{(0)}_\pi-\eps(q)}\:.\label{eq:kp}
\end{equation}
Here, the prime means the differentiation over $K$ at the point $K=\pi$.
The first term, $\propto \langle 0|\mathcal H''|0\rangle$ comes from a linear perturbation theory and the second term describes an admixture of the scattering states Eq.~\eqref{eq:scat} to the bound state Eq.~\eqref{eq:bound} for $K\ne \pi$.
The first term in Eq.~\eqref{eq:kp} is easily found after direct substitution of the wave function Eq.~\eqref{eq:bound} as $\langle 0|\mathcal H''|0\rangle= 4\Gamma_0\cos 2\phi  (2-\cos^2 2\phi)/\sin^3 2\phi$. 
The calculation of the second term in Eq.~\eqref{eq:kp} is also straightforward but quite tedious. However, the overall result is surprisingly compact,
\begin{equation}\label{eq:mass}
\frac1{m}=-\frac{\Gamma_0\sin\varphi \cos 3\phi}{8\cos^6\varphi}\:.
\end{equation}
Equation Eq.~\eqref{eq:mass} demonstrates, that for $\varphi=\pi/6$, when $\cos3\varphi=\pi/2$, the effective mass diverges, $1/m= 0$. The
analytical result Eq.~\eqref{eq:mass}, shown by thin black line in Fig.~\ref{fig:2}(a) is in perfect agreement with the  direct numerical calculation (black dots). 

The divergence of the effective mass at the ``magic'' distance $d=\lambda_0/12$ does not seem to have a trivial explanation, at least within the framework of $\bm k\cdot \bm p$ perturbation theory. It apparently results from a delicate balance between the first and second terms in Eq.~\eqref{eq:kp} for the inverse mass. The contributions of these terms are separately shown by blue and red lines in Fig.~\ref{fig:2}(b), and their sum is given by a black line. The large mass of the bound pairs also seems to be related to the fact that the bound state branch is located between the upper and lower polariton bands with small and large relative motion wave vectors $q$, respectively. As such, the interaction with upper and lower polariton bands pushes the bound pair band to the  opposite directions, effectively  making it flatter.

The lifetime of photon pair in the finite array is shown by the green curve in Fig.~\ref{fig:2}(a). In accordance with the results of Zhang et al. in Ref.~\cite{Zhang2019arXiv},  the lifetime has a sharp maximum for  $d=\lambda_0/12$. Being armed with the analysis of the dispersion we can now provide a very crude explanation of this lifetime enhancement. Namely, the bound photon pair band becomes quasi-flat due to the diverging effective mass. Hence, the bound pair can not reach the edge of the structure and can not radiate.  More microscopic details of this effect are given in the following Section.

\section{Bound pair near the edge}

In this section we outline a    mechanism to explain how exactly the diverging effective mass of the bound photon pair quenches its radiative escape and boosts the lifetime. While the final result is very intuitive, the microscopic details are rather untrivial and involve an intricate interplay between the interaction potential, binding the pair, and the confining potential at the edge of the structure.

It is very instructive to simplify the problem by rewriting the two-particle Schr\"odinger equation Eq.~\eqref{eq:Sch} for the wavefunction \cite{Zhong2019}
\begin{equation}\label{eq:chi}
\chi=H_0^{-1}\Psi H_0^{-1}\:,
\end{equation}
rather than the original two-particle wavefunction $\Psi$. The reason is that the matrix 
\begin{align}\label{eq:iH}
&H_0^{-1}\Gamma_0\\=&\begin{pmatrix}
 -\frac{1}{2}\cot \phi+\frac{\rmi}{2}&\frac1{2\sin\phi}&0&\ldots\\
 \frac1{2\sin\phi}&-\cot \phi&\frac1{2\sin\phi}&\ldots\\
0&  \frac1{2\sin\phi}&-\cot \phi&\frac1{2\sin\phi}\ldots\\
&&\ddots&\\
  \ldots& \frac1{2\sin\phi}&-\cot \phi& \frac1{2\sin\phi}\\
 \ldots&0& \frac1{2\sin\phi}& -\frac{1}{2}\cot \phi+\frac{\rmi}{2}
 \end{pmatrix}\nonumber
  \end{align}
 is just a three-diagonal one.   As such, the transformation Eq.~\eqref{eq:chi} allows one to get rid of the long-range interaction inherent to the original dense single particle Hamiltonian $H_0$ in Eq.~\eqref{eq:H}. We stress that {\it  the imaginary part of the matrix $H_0^{-1}$ is nonzero only at the edges of the array}. This is quite intuitive,  since the  radiative losses in the considered waveguide are possible only via the photon escape through the edges.  The transformed Schr\"odinger equation Eq.~\eqref{eq:Schb} reads
 \begin{equation}
 H_0^{-1}\chi+ \chi H_0^{-1}-2\diag [ (\diag \chi H_0^{-1})]=2\eps H_0^{-1}\chi H_0^{-1}
  \end{equation}
and involves only sparse  three-diagonal matrices. As such, it is much easier to analyze than the original Eq.~\eqref{eq:Schb}. Moreover, in the case when $\varphi\ll 1$ the operator $H_0^{-1}$ becomes just a second discrete derivative operator, $ H_0^{-1}\approx \Gamma_{0}/(2\varphi) \partial^{2}$~\cite{Zhong2019}.
  The radiative decay rate of the two-photon state can be presented as  \cite{Ke2019arXiv}
   \begin{equation}
 -\Im \eps=\Gamma_{0}\sum\limits_r|d_r|^2,\quad  d_r=\sum\limits_{s=1}^N\Psi_{rs}\e^{\rmi \varphi s}\:.
 \end{equation} 
For the transformed wave function we find  $d_r\propto [H_0^{-1}\chi]_{r1}$. Given that the matrix $H_0^{-1}$ is three-diagonal, the radiative decay rate is expressed only via the values of the
 new two-photon wavefunction $\chi$ only at the edges, 
 \begin{multline}\label{eq:imeps1}
-\Im \eps=\sum\limits_{r=1}^{N} \left|[H_0^{-1}\chi]_{r1}\right|^2
\approx \Gamma_{0}\sum\limits_{j=2}^{N-1} |d_r^2|, \\\:  |d_r^2|=\frac{|\chi_{j+1,1}+\chi_{j-1,1}-2\cos\phi \chi_{j,1}|^2}{4\sin^2\phi}.
\end{multline}
Here we have neglected for simplicity the contribution from the points $j=1,N$ where the  matrix elements of the Hamiltonian Eq.~\eqref{eq:iH} differ from their bulk values which is a reasonable approximation for large number of atoms $N\gg 1 $.

\begin{figure}[t!]
\centering\includegraphics[width=0.48\textwidth]{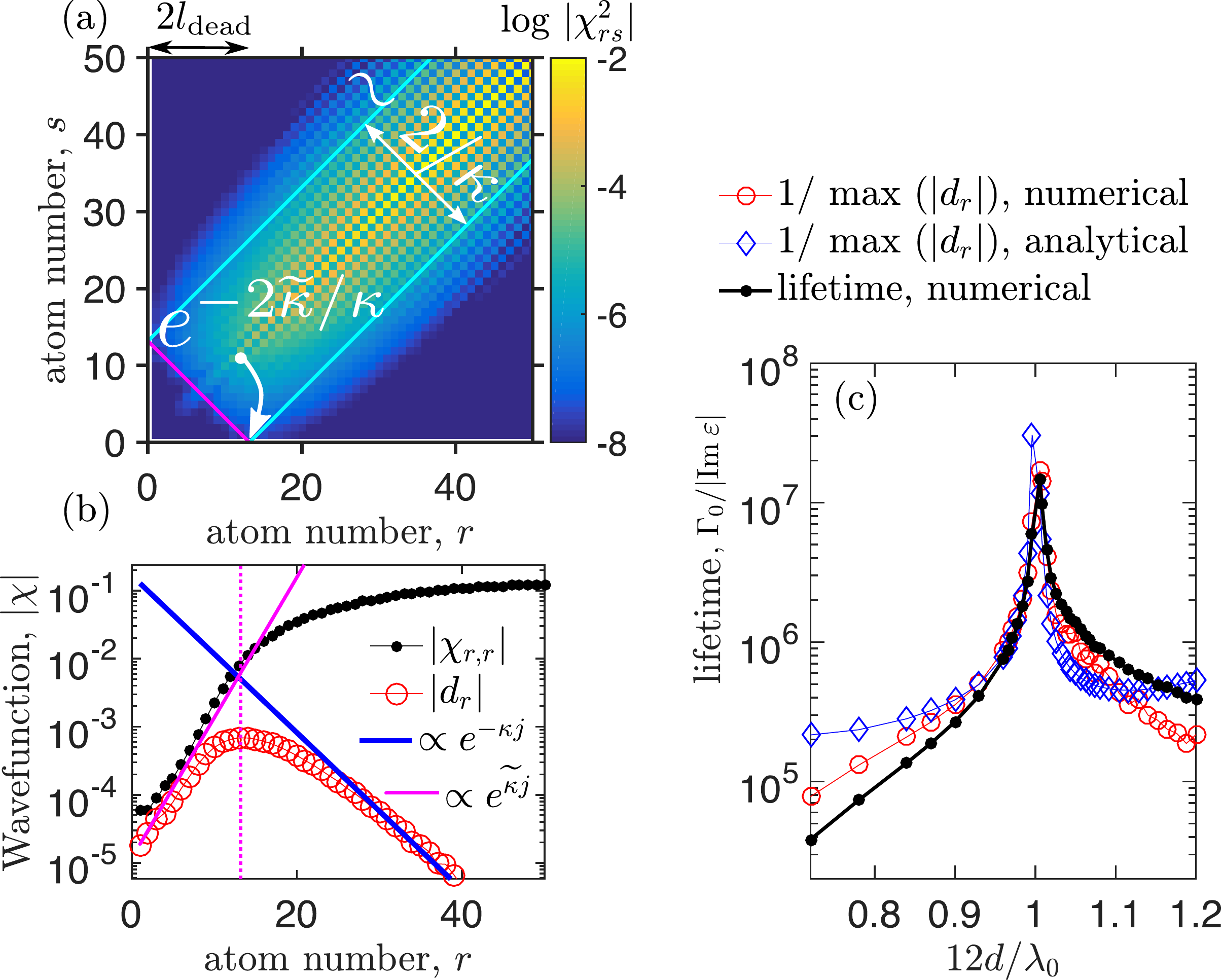}
\caption{(a) Spatial map of the wave function of the confined two-photon state $\log |\chi_{rs}|^2$ near the edge of the array. The state has the eigenenergy $\eps/\Gamma_0\approx 1.45-3.73\times 10^{-6}\rmi  $ and is calculated  
for $12d/\lambda_0=0.9$ and $N=100$~atoms. Magenta line  drawn in the direction $r=-s $ shows the dead layer position. Cyan lines 
along the direction $r=s $ illustrate the bound pair size. The lines cross the edges at  the points 
 $r\equiv 2l_{\rm dead}=7/\kappa$, $s=0$ and $s=2l_{\rm dead}$, $r=0$. (b) Black solid dots: diagonal cross-section of the two-photon state $\chi_{rr}$ in panel (a).
Red open circles: distribution of the effective radiation amplitudes $|d_r|$ from Eq.~\eqref{eq:imeps1}.
Vertical dotted magenta line is drawn at $r=2l_{\rm dead}$. Magenta  and blue lines show the asymptotic expressions with the exponents 
Eq.~\eqref{eq:varkappa} and Eq.~\eqref{eq:kappa}, respectively. (c)  Black dots: lifetime of the bound photon pair as a function of the array period $d$. 
Open red circles: maximum value of the amplitudes $|d_r|$ as functions of $r$ plotted for different array periods $12d/\lambda_0$. Blue  diamonds: analytical asymptotic expression 
Eq.~\eqref{eq:an}.
 }
\label{fig:3}
\end{figure}

Equation Eq.~\eqref{eq:imeps1} for the radiative decay rate provides a rigorous  foundation for the following analysis. We will now numerically demonstrate that the values of $\chi_{j,1}$ 
at the boundary are sensitive to the translational mass of the bound pair $m$ which in turn affects the radiative decay.
In Fig.~\ref{fig:3}(a,b) we examine the spatial distribution of the confined wave function $\chi$ near the edges of the structure. The overall map of the distribution for the period $d=0.9\lambda_0/12$ is presented in Fig.~\ref{fig:3}(a). The bound pair has size on the order of 
$1/\kappa$ in the direction $r=-s$ (cyan lines in Fig.~\ref{fig:4}a), transverse to the center of mass motion direction $r=s$. Next, we analyze the dependence of the wave function on the center of mass coordinate $(r+s)/2$, i.e.  the diagonal cross section $\chi_{rr}$ of the distribution in Fig.~\ref{fig:3}(a). The result, shown by the black dots in Fig.~\ref{fig:3}(b),  does not decay to the edge as a linear function of the distance, as would be expected from the infinite wall boundary condition when $\chi_{rr}\propto \sin(\pi r/N) $. Instead, the center of mass motion is  strongly  suppressed already when the  distance is below a certain threshold $l_{\rm dead}$.  This effect 
is  well known in the physics of Wannier-Mott excitons, bound electron-hole pairs, confined in semiconductor nanostructures. It can be phenomenologically described as a formation of a so-called ``dead layer''  with the thickness on the order of the exciton Bohr radius~\cite{Hopfield1963,Excitons,Ivchenko2019}. As a result, the effective barrier for excitons  should be placed not exactly at the edge but at a certain distance.  This is illustrated by the magenta line in Fig.~\ref{fig:3}(a). For  distances closer to  the edge than the dead layer thickness the wave function decays exponentially rather than linearly.
  Numerical analysis of the wavefunction for $N=100$ shows that in the range of $0.6\le 12d/\lambda _0\le  1.4$ this thickness can be satisfactory approximated by $ l_{\rm dead}\approx 3.5/\kappa$.

 Red dots in Fig.~\ref{fig:3}(b) show the spatial distribution of the distribution $|d_r^2|$ from Eq.~\eqref{eq:imeps1} for the eigenstate in Fig.~\ref{fig:4}(b). The overall radiative decay rate is given by the integral of this  distribution. Crucially, the distribution has a distinct maximum at the distance on the order of  $ 2l_{\rm dead}\approx 7/\kappa$ from the edge. Hence, the radiative lifetime is determined by the probability of one particle to be exactly at the edge and the second one to be close to the edge at the same time. The spatial distribution of the  amplitudes $d_r$ is bounded by two asymptotic  exponential functions The first one, shown by the blue line in Fig.~\ref{fig:3}(b), decays to the bulk at the  exciton size $\propto \exp(-\kappa j)$. The second one decays as $\exp(\varkappa j)$ to the edge with the   exponent
\begin{equation}\widetilde\kappa=\sqrt{2mU},\label{eq:varkappa}
\end{equation}
where  $U\approx 0.002$ for $N=100$ is the numerical fit parameter weakly depending on $12d/\lambda_0$. Physically, Eq.~\eqref{eq:varkappa} describes effective tunnelling of the bound pair under the phenomenological  barrier of the height $U$ that is present  at the distances from the edge shorter than the dead layer thickness. Our numerical analysis shows that the barrier height $U$ is practically independent of  the array period in considered range of $0.6\le 12d/\lambda _0\le  1.4$. The decay parameter Eq.~\eqref{eq:varkappa} depends on the effective mass of the bound pair $m$ and hence is very large for $12d/\lambda
\approx 1$. The position of the maximum of the amplitudes $|d_r|$ as a function of the array period is close to $2l_{\rm dead}$ (dotted magenta line in Fig.~\ref{fig:3}b). Next, in Fig.~\ref{fig:3}(c) we demonstrate that the lifetime and  the maximum of $|d_r|$ scale in the same way with the variation of the period of the array. The whole sum over different points $r$ in Eq.~\eqref{eq:imeps1} is mostly determined by the maximum value of $|d_r|^2$ and is not dependent on the width of the distribution $|d_r^2|$. This maximum value can be estimated by comparing  the wavefunction at the diagonal (black points in  Fig.~\ref{fig:3}b) and at the edge (red circles in Fig.~\ref{fig:3}b) for $r=2l_{\rm dead}$. It turns out that the value at the edge is suppressed by the parameter on the order of $\e^{-\widetilde \kappa/\kappa}$. Physically, it means that in {\it order to reach the edge the bound pair has to tunnel as a whole by the distance on the order of its  size}. This tunneling process is illustrated by the curved white arrow in Fig.~\ref{fig:3}(a). As such, the maximum value of $|d_r|^2$ at the boundary can be approximated by
$\max |d_r^2| \propto  |\chi_{2l_{\rm dead},2l_{\rm dead}}^2| \e^{-2\widetilde \kappa/\kappa}$. Given that for a standing wave with 
$\chi_{rr}\propto \sin(\pi r/N) $ the value of $\chi_{2l_{\rm dead},2l_{\rm dead}}$ is proportional to $l_{\rm dead}$, we arrive at the asymptotic expression 
\begin{equation}\label{eq:an}
\max |d_r^2|   \propto l_{\rm dead}^2\e^{-2\widetilde \kappa/\kappa}
\end{equation}
for the radiative decay rate.
The result Eq.~\eqref{eq:an} is plotted by the blue diamonds in Fig.~\ref{fig:3}(c) and satisfactory describes the behaviour of the lifetime as a function of array period and its enhancement  at $12d/\lambda_0=1$. In this case the effective mass diverges. Thus, one has $\widetilde \kappa \gg 1$ so that the pair tunneling $\propto \e^{-2\widetilde \kappa/\kappa}$ is suppressed, and the radiation  of bound photons through the edge is strongly quenched.

\section{Lifetime oscillations with the number of atoms}
\begin{figure}[t!]
\centering\includegraphics[width=0.48\textwidth]{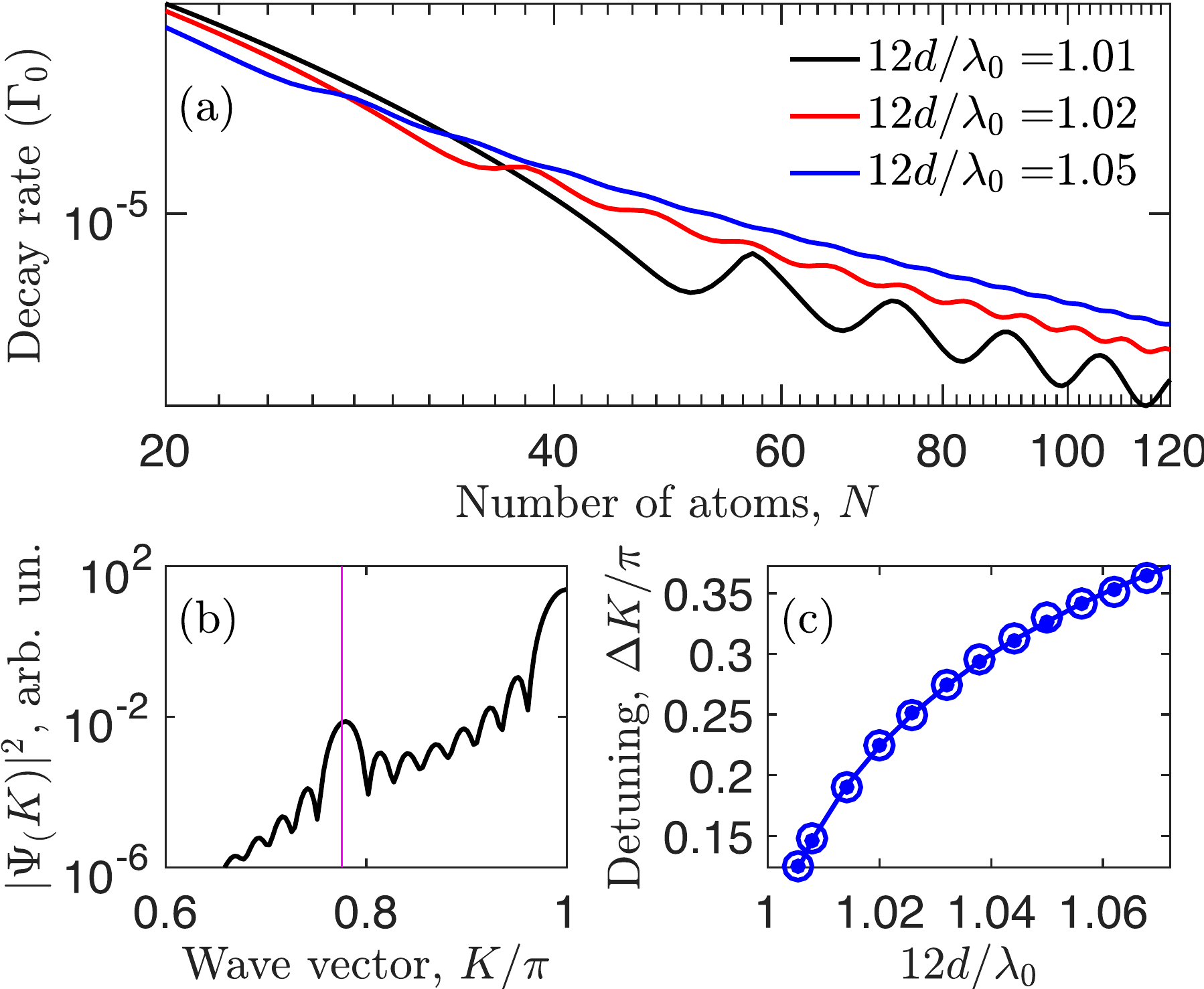}
\caption{
(a) Radiative decay rate of the most subradiant bound two-photon pair states depending on the number of atoms in the array $N$. Calculated for 3 different periods slightly above $\lambda_0/12$, indicated on graph. 
(b) Fourier transformation of the wave function $\Psi_{r,r+2}$. Vertical line shows the point $K-\Delta K $, extracted from the bulk dispersion.
Calculated for $N=100$ atoms and $12d/\lambda_0=1.02$.
(c). Circles show the wave vector $\Delta K$ extracted from the peak position of the Fourier transform  of the lifetime dependence on $N$ as a function of the array period. Lines show the dependence of the wave vector $\Delta K$ corresponding to the point where $\eps_{\pi-\Delta K}^{(0)}=\eps_{\pi}^{(0)}$, see Fig.~\ref{fig:1}(a).
 }
\label{fig:4}
\end{figure}

We will now examine one more puzzling feature of the ``magic period'' $d=\lambda_0/12$, revealed in Ref.~\cite{Zhang2019arXiv}.
Namely, for $d$ slightly larger than $\lambda_0/12$ the dependence of the radiative decay rate of the confined bound pairs on the number of atoms in the array becomes nonmonotonous. This is also demonstrated by our calculation in Fig.~\ref{fig:4}(a) showing the radiative decay for three different periods.  The oscillations are quite pronounced for $12d/\lambda_0=1.01$ (black curve) and then fade away when the period increases to $12d/\lambda_0=1.05$ (blue curve). We will now demonstrate, that these oscillations are a direct consequence of the nonmonotonous dispersion of the bound pairs in the vicinity of the point $K=\pi$ for $d>\lambda_0/12$, see Fig.~\ref{fig:1}(b).

The dispersion curve for $d>\lambda_0/12$ can be  approximately presented as
\begin{equation}
\eps^{(0)}_{K}\approx \eps^{(0)}_{\pi}-\alpha (K-\pi)^4+\frac{(K-\pi)^2}{2m}
\end{equation}
where $\alpha>0$ and $|K-\pi|\ll \pi$. Clearly, the dispersion in the bulk is degenerate. Namely, there exist two points with the same energy $\eps^{(0)}_{\pi}$,
$K=\pi$ and $K=\pi-\Delta K$, where 
\begin{equation}
\Delta K=\sqrt{2m\alpha}\:.\label{eq:DK}
\end{equation} In the infinite structure the two states with the center-of-mass wave vectors
$\pi$ and $\pi-\Delta K$ are independent because of the momentum conservation law. However, in the finite array the translational symmetry is broken at the boundary. The  center-of-mass wave vector is no longer conserved exactly and the mixing between the confined states with different center-of-mass wave vectors becomes possible. This effect is  well known for confined states in semiconductor nanostructures. The especially relevant case is presented by silicon quantum wells~\cite{nestoklon1}. In silicon the conduction band dispersion along the [001] direction is nonmonotonous  and the band extremum is shifted from the edge of the Brillouin zone by approximately 15\% to the point $K=0.85\pi/a$ ($a$ is the lattice constant). Surprisingly, the electron dispersion in silicon looks very much like the red curve in Fig.~\ref{fig:1}(b) turned upside down. In silicon quantum wells, grown along the [001] direction, the states corresponding to the opposite valleys $K=0.85\pi/a$ and $K=-0.85\pi/a$ are split, and the magnitude of the valley-orbit splitting oscillates as a function of the quantum well thickness with the wave vector $2\Delta K=2(\pi/a-K)$.

 A very similar effect happens in the considered quantum optical system. Namely, the most subradiant  confined pair states correspond mostly to the bulk wave vector $K=\pi$. However, in the finite structure the confined state acquires a small admixture of the states with small center of mass wave vector $K-\Delta K$. This can be directly seen by analysing the spatial distribution of the wave function of the confined pair. Namely, we perform the Fourier transformation for the confined pair wave function $\Psi_{r,r+2}$ along the center-of-mass direction $r$.  The Fourier transformation reveals has two peaks, shown in Fig.~\ref{fig:4}(b). The main peak is at $K=\pi$, but  there also exists a second peak at $K=\pi-\delta K\approx 0.78\pi$, shown by the magenta line. 
 The magnitude of the admixture of the states with the center-of-mass wave vector $\pi-\delta K$    is proportional to the Fourier component of the structure factor with the wave vector $\Delta K$~\cite{nestoklon1,Poddubny2012}. As such, the admixture and the complex energy of the eigenstate will have a contribution oscillating with the number of atoms.  The fading of oscillations for larger periods, seen in Fig.~\ref{fig:4}(a), is also clear. In this case the wave vector $\Delta K$ becomes larger. As such, it gets harder to couple the states with the center-of-mass momenta $K=\pi$ and $K=\pi-\Delta K$ and the interference effects are washed out.

 In order to further confirm this hypothesis we have extracted the period  of the oscillations by performing the Fourier transform of the lifetime dependences in Fig.~\ref{fig:4}(a) for different array periods. Next, in Fig.~\ref{fig:4}(c) we compare the obtained wave vectors $\Delta K$ (blue dots) with the wave vector   Eq.~\eqref{eq:DK} found from the bound pair dispersion in the infinite structure (blue solid lines). The results are in perfect agreement, confirming our explanation of the lifetime oscillations. 
 
%

\section{Conclusions and Outlook}
To conclude, I have analyzed in detail the dispersion of bound photon pairs in a periodic array of two-level atoms in a waveguide. By means of numerical calculations and analytical $\bm k\cdot \bm p$ perturbation theory I have demonstrated that the sign of the dispersion curvature at the Brillouin zone edge changes at the ``magic'' array period $d=\lambda_0/12$, corresponding  to the quasi-flat band of the bound pairs. 
 Due to the uncovered transformation of the two-photon Hamiltonian to the  nearest-neighbor coupling model,
 the radiative decay has been rigorously linked to the values of the wave function  only at the edges  of the array. Next, I have shown how the diverging mass of bound pairs  suppresses the photon amplitude  at the edges and demonstrated a  connection of this effect to the quenched radiative losses.  These results provide a qualitative and quantitative explanation of the unusually long radiative lifetime of the bound photon pairs for $d=\lambda_0/12$~\cite{Zhang2019arXiv}: the  pair becomes heavy and can not escape the structure. 

I also demonstrate that for the array period slightly larger than $\lambda_0/12$ the dispersion curve is nonmonotonous near the Brillouin zone edge. This results in a non-monotonous dependence of the radiative decay rate on the number of atoms in a finite array. The radiative decay rate oscillations  are surprisingly similar to the oscillations of valley-orbit and spin-orbit splittings, predicted  for semiconductor nanostructures, such as silicon quantum wells~\cite{nestoklon1}. 

These findings indicate that despite  almost 70-year history, the Dicke-like models remain full of surprises even when constrained by the rotating wave approximation, Markovian approximation and one spatial dimension. The propagation and radiative decay  of many-body collective light-coupled excitations  merit further investigation.  Further  unexpected and  fruitful cross-disciplinary links between the few-body quantum optics and the physics of excitons in semiconductor nanostructures are possible. Moreover,  the quasi-flat bands with high density of bound pair states might be useful to enhance the nonlinear interactions~\cite{Leykam2013,Ke2019arXiv}. It is also very interesting to examine how  the concept of bound states in continuum, being now actively developed in classical optics~\cite{Hsu2016}, can be generalized to the many-body interacting quantum regime  and whether it could be used in practice to prolong the life of quantum correlations.

\begin{acknowledgments}

I am deeply grateful to K. M\o lmer and Yu.-X.~Zhang  who attracted my interest to this problem.
I acknowledge useful discussions with L.E. Golub, E.L.~Ivchenko, Y.~Ke, K.~Koshelev, A.V.~Poshakinskiy, M.A.~Semina and J. Zhong.
\end{acknowledgments}

%

\end{document}